# MoSe$_2$-enhanced novel polyacrylamide composites with tunable refractive index and band gap energy

**Bengü Özuğur Uysal[a] *, Önder Pekcan[a]**

[a] Kadir Has University, Faculty of Engineering and Natural Sciences, Cibali, Fatih, Istanbul 34083, Turkey

 * *Corresponding author*

*Tel: +90 212 533 65 32 / 1345*

*Fax: +90 212 533 65 15*

*e-mail addresses:*

*bozugur@khas.edu.tr*

*pekcan@khas.edu.tr*

**Abstract**

Hydrogel/inorganic composites have attracted attention in many applications, such as opto-electronic devices, biosensors, catalysis, and energy storage because of their capacity to increase and regulate optical and electronic features. In this study, MoSe$_2$ enhanced polyacrylamide composites were formed using free radical crosslinking copolymerization process. The uv-vis spectrometer was used to investigate the amount dependent optical properties of new composites. The absorbance, transmittance, band gap energy, extinction coefficient, and refractive index of composites were investigated in detail. It has been found that as the amount of MoSe$_2$ increases, the band gap energy value decreases, which proves that the material has higher absorption and becomes more conductive. In addition, the optimum amount of MoSe$_2$, which has a very high optical transmittance in the visible region and also has a high conductivity, has been determined. The results reveal that a promising composite has been produced for many solar energy applications, including optoelectronic applications.

## 1. Introduction

Recent research interests are on polymer composites, including both organic and inorganic components such as metal chalcogenides and hydrogels [1]. They are one of the most extensively studied materials due to their outstanding electrical, optical and mechanical properties. Based on optical materials such as molybdenum selenide ($MoSe_2$) from the family of metal chalcogenides, several reports presented for thermoelectric, photodetector pulse-laser and photonics applications [2-6], because of its remarkable optoelectronic properties. According to Beal et. al [7], at wavelength of 650 nm, optical constants of $MoSe_2$ such as refractive index, n = 4.7495; extinction coefficient, k=1.1893; dielectric constants: $\epsilon_1$=21.143, $\epsilon_2$=11.297 and absorption coefficient, $\alpha$ = 2.2993 x $10^5$ $cm^{-1}$ has been derived precisely. The indirect bandgap of $MoSe_2$ in its bulk-form has a value of 1.1 eV. A direct band gap of 1.5 eV can be obtained by exfoliating $MoSe_2$ into a few layers [8]. In particular, the development of the photoelectron $MoSe_2$ nanoflower structure marked the beginning of the investigation into the potential of photodynamic therapy, as the electrons and holes on the surface of semiconductors with positive valence band potential and negative conduction band potential could potentially react with the surrounding reactants to form reactive oxygen species [9]. Therefore, $MoSe_2$ can be considered as a promising electroactive material, and it is undeniable that it has the ability to demonstrate electrocatalytic and photocatalytic performance in solar cells [10] and other related applications [11-13]. Furthermore, when incorporated together with polymers, the resulting composite or hybrid material may also have other superior properties [14-17]. For instance, $MoSe_2$ based chemical sensor or biosensor can be obtained [18]. In the presence of a capping agent, PEG, a hybrid material with $MoSe_2$ nanodots fixed on $MoSe_2$ nanosheets, was synthesized by one-pot synthesis [19]. PVP has been applied to exfoliating $MoSe_2$ nanosheets to get higher photothermal conversion efficiency, resulting in the reduced viability of Hela cells [20]. Furthermore, the roles of $MoSe_2$-nanocomposites in diagnosis and therapy of cancer have been presented [21]. Recently, increased Li-ion rate capability and stable efficiency has been made possible by $MoSe_2$ nanosheets in silicon oxide fiber electrodes made of polymers [22]. Memory device performance can be improved utilizing $MoSe_2$ and PVA [23]. On the other hand, polyacrylamide (PAAm), the most common hydrogel, is inexpensive, can be cast into desired shapes, and has flexibility for many applications. The refractive index of PAAm has been found to vary within the range of 1.4 to 1.5. Therefore, in this work, the amount dependent optical properties of $MoSe_2$ enhanced PAAm were investigated. It is normally accepted that composite hydrogels can effectively connect the unusual properties of the inorganic ($MoSe_2$) and organic (PAAm) components to get a desired effect.

## 2. Materials and Methods

*2.1. Preparation of MoSe$_2$ enhanced PAAm Gels*

Acrylamide (AAm) and crosslinker N'–Methylenebisacrylamide (Bis) monomers and the initiator ammonium persulfate (APS) were dissolved in distilled water by magnetically stirring. Tetramethyl ethylene diamine (TEMED) was added as an accelerator. Polyvinyl pyrrolidone (PVP) was added to the solution. This preparation method is given in our previous studies with different amounts of BIS, PVP, MoS$_2$, and WS$_2$ [24-26]. Distilled water prepared by KHAS Chemistry Laboratory was used as the reaction solvent. MoSe$_2$-enhanced PAAm gels have been obtained using free radical crosslinking copolymerization technique after addition of Molybdenum selenide (MoSe$_2$). The concentrations of AAm, Bis, APS, PVP and TEMED were taken to be constant throughout the polymerization process, and only the MoSe$_2$ amount varied from 0.5 to 2.5 mg. The gelation process of AAm is quite fast. Therefore, immediately after the reaction catalyst TEMED was added to the solution, half of the solution was left in the beaker and the other half was poured into the quartz cuvette and quickly placed in the spectrophotometer cabinet. Of the hard gels formed after drying in the open air under laboratory conditions, those that are dried in a quartz cuvette have the shape of a square prism and those that are dried in a beaker have a disc shape (Figure 1). Figure 1 c and d depicts the disc shaped samples after drying in air.

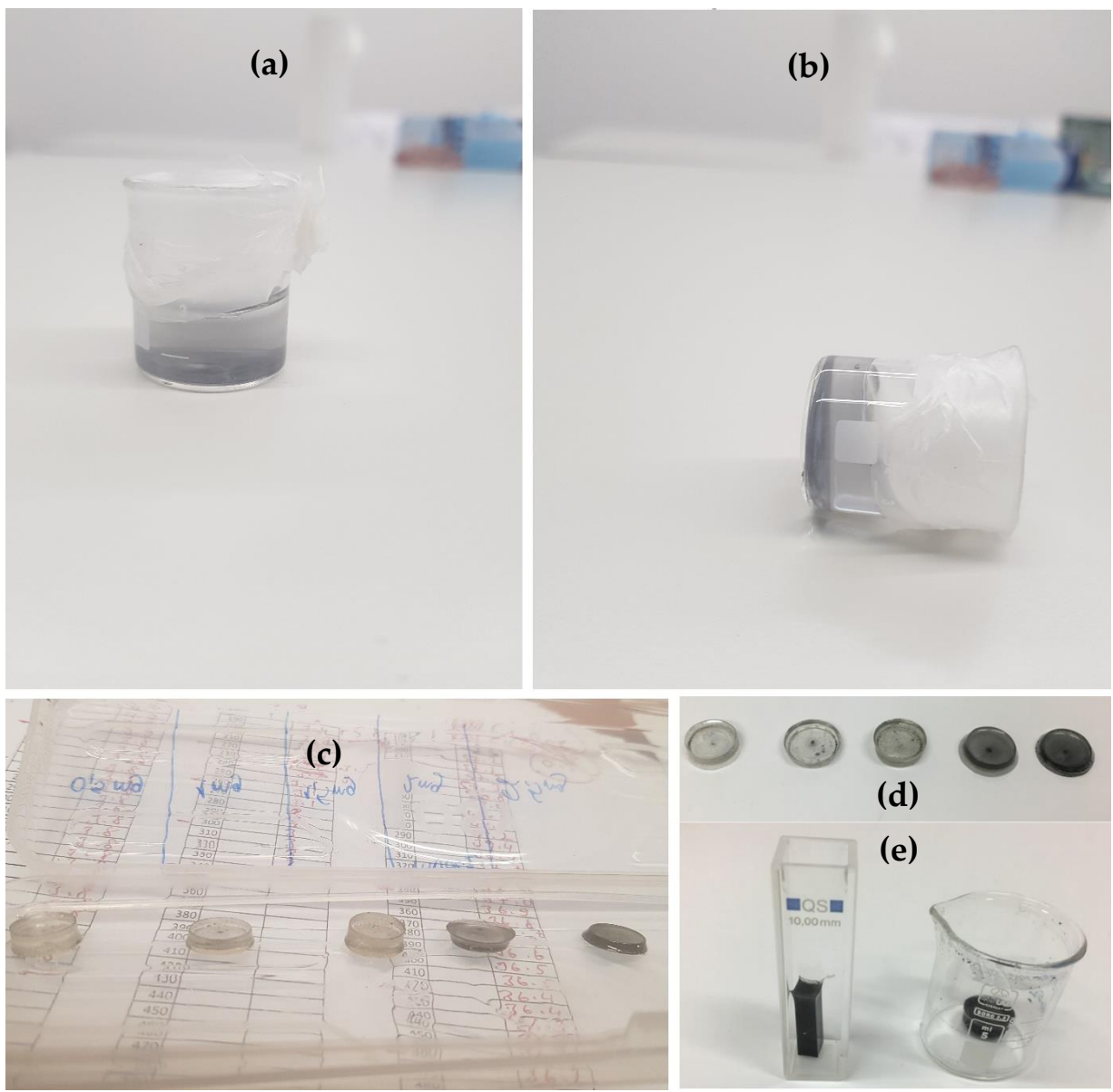

**Figure 1. (a)** Hydrogel formation in the beaker. **(b)** Transparent and non-fluid form of gel. **(c), (d)** Hydrogels with different amounts of MoSe₂ after drying in air. Discs are ready for absorption measurements. From left to right, 0.5 mg to 2.5 mg. **(e)** Hydrogel in the form of square prism and disc after drying in air.

*2.2. Characterization*

The composite gels with various amounts of MoSe$_2$ were measured using Labomed Spectro 22 uv-vis Spectrophotometer at the wavelength range from 190 to 1100 nm. In the graphs given, the data in the wavelength intervals were evaluated in the walnut intervals, which are thought to be theoretically linked to a certain model. Absorbance and transmittance measurements were taken for single wavelength one by one. Absorbance and transmittance were noted after the calibration of the spectrophotometer for each wavelength. All hydrogels are allowed to dry at room temperature. After drying, the change of the absorption coefficient with various amount of MoSe$_2$ is measured at $\lambda$=700 nm wavelength. Transmission values are measured with different wavelength, and then band gap energy is calculated from the Tauc's plot by varying absorbance values with respect to wavelength. Then, extinction coefficient, and refractive index of composites were investigated in detail.

## 3. Results and Discussion

*3.1. Transmittance and absorbance response of composite gels*

After the drying process at room temperature, transmittance and absorbance response (Figures 2 and 3) of MoSe$_2$ – PAAm composites with various MoSe$_2$ content are measured using uv-vis spectrometer at different wavelengths. The longer the wavelengths, the greater the transmittance of the MoSe$_2$-enhanced PAAm composites in visible region (Figure 2). Moreover, it was found that the transmittance value is decreasing with MoSe$_2$ amount inside composite gel. In Figure 3, the intensity of the absorbance peaks does not change much for the composite gels prepared with different MoSe$_2$ amounts. It can be thought that the absorbance values do not change despite the serious change in the transmittance values due to the scattering effect that may be caused by the distribution of MoSe$_2$ nanosheets in polyacrylamide. As it is known, when light interacts with any substance, it is expected to pass through the substance, be absorbed into the substance, or be reflected by the substance. However, the fact that there are different structures in matter beyond its nature, and even the size of these structures is comparable to the wavelength of light, brings about the phenomenon of scattering. On the other hand, Figure 3 represents the absorption edge of the composite gels. The absorption edge is observed to be red shifted for composite gels with higher amount of MoSe$_2$.

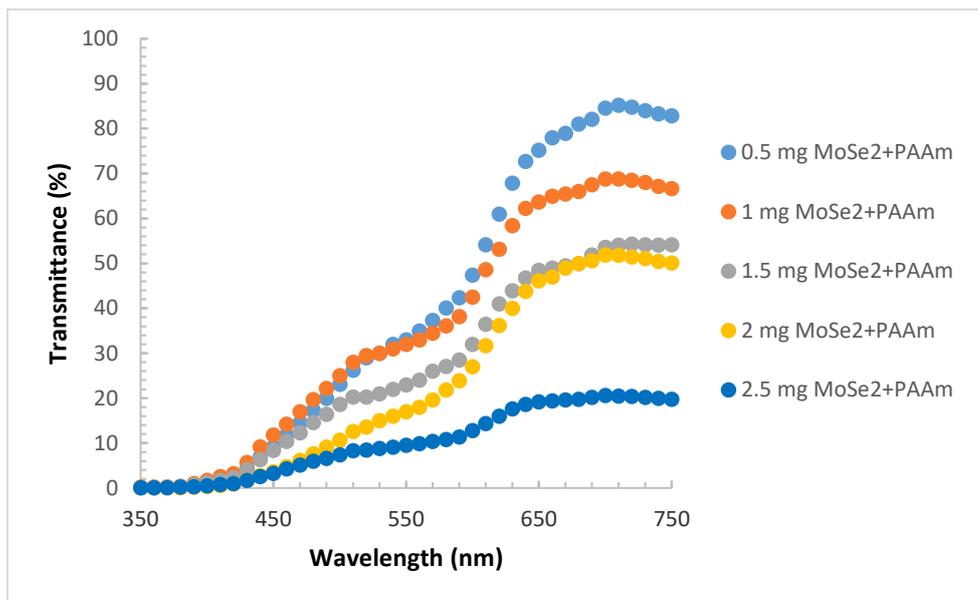

**Figure 2.** Wavelength dependence of the transmittance of MoSe$_2$ enhanced PAAm composites.

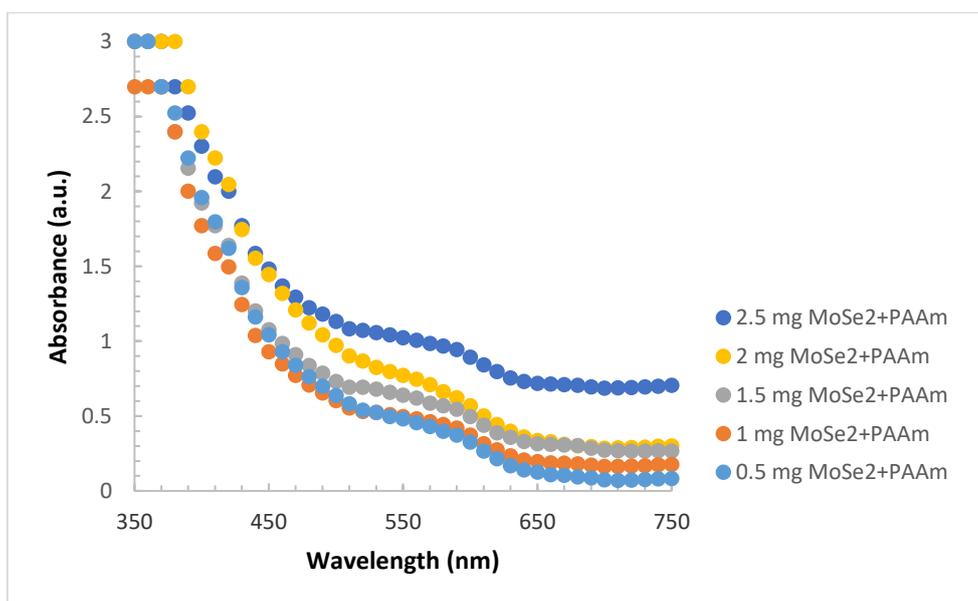

**Figure 3.** Absorbance of MoSe$_2$ enhanced PAAm composite gels with various MoSe$_2$ content.

*3.2. Band gap energy determination of composite gels*

Absorption coefficient, $\alpha$ values are obtained by the formula:

$$\alpha(h\nu)=(\log_e 10/t)\, A(h\nu) \tag{1}$$

where t is the thickness of the discs, and A is the absorbance values for corresponding photon energies (hv). The band gap energies of the composite gels, $E_g$ are found using the following Tauc's formula for direct transition [27].

$$(\alpha h\nu)^2 = C(h\nu - E_g) \qquad (2)$$

As seen in Figure 4, after obtaining $(\alpha h\nu)^2$ values for each dry hydrogel, the value of crossing photon energy is taken as the band gap energy. We obtained band gap energy by extrapolating of $(\alpha h\nu)^2$ linear fit. Band gap energy is plotted versus the amount of $MoSe_2$ content in Figure 5. It was found that the band gap decreases with increasing the values of $MoSe_2$ content and then increases with further increases (Figure 6). The transmittance and band gap energy values of the composite gels are listed in Table 1. It will not be enough for these gels to just be transparent, conductive, or elastic if they are to be employed for a multifunctional application. Those had an average transmittance value that may be used to create a composite gel that combines as many qualities as possible. Therefore, these gels should be discarded if their transmittance values are 20% or lower at most wavelengths in the visible region. Composite gels containing 0.5 and 1 mg $MoSe_2$ can be considered transparent according to the transmittance values at 700 nm. On the other hand, band gap energies do not vary much for different gels with the amount of $MoSe_2$. The values of band gap energy have been found to be in the range of 2.63– 2.74 eV. Since these values are in the energy value range required for many applications such as biomedical, semiconductor, transparent conductor, photosensor [28-30]. By changing the amount of $MoSe_2$, band gap energy can be regulated in accordance with the desired optical and electronic properties for multifunctional applications. Using this information, another group of researchers can easily obtain the composite gel by determining a proper amount of $MoSe_2$ according to the desired band gap energy and transmittance combination for their specific application. In addition to these, mechanical properties must also be taken into consideration. It has also been theoretically proven that $MoSe_2$ has very good mechanical properties [31]. Therefore, as the amount of $MoSe_2$ integrated into the already flexible polyacrylamide gel increases, the mechanical properties of the resulting composite will also improve. This has been shown in previous studies for composites formed by polyacrylamide with other metal dichalcogenide group materials [24, 25]. Given all of these findings, it is projected that polyacrylamide doped with 0.5–1 mg $MoSe_2$ may produce the best outcomes.

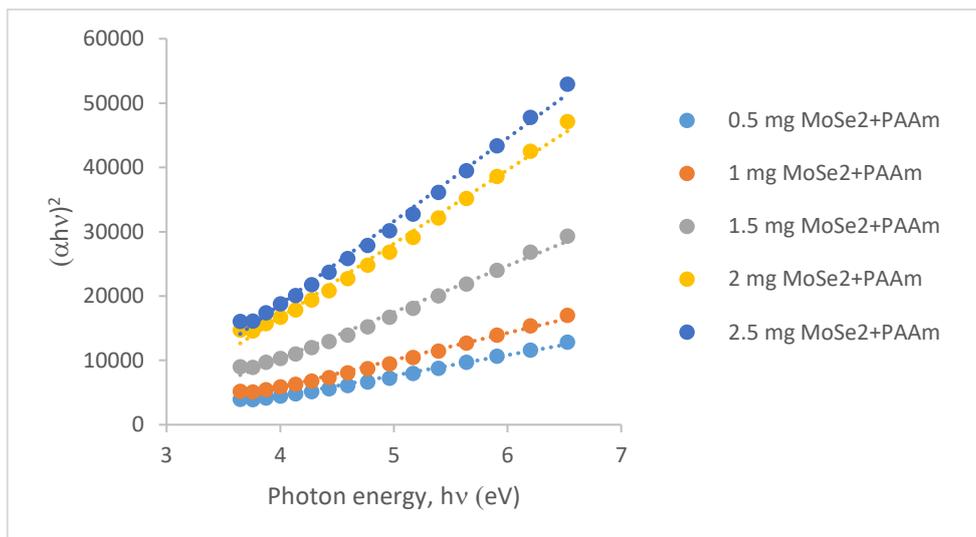

**Figure 4.** ($\alpha h\nu)^2$ versus photon energy of MoSe$_2$ enhanced PAAm composite gels.

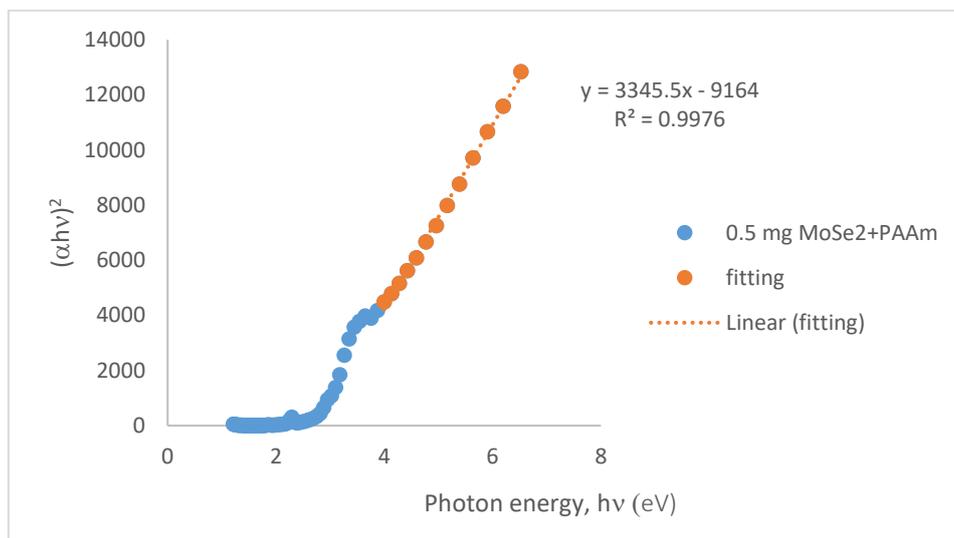

**Figure 5.** Extrapolation of versus photon energy for 0.5 mg of MoSe$_2$ enhanced PAAm composite gels.

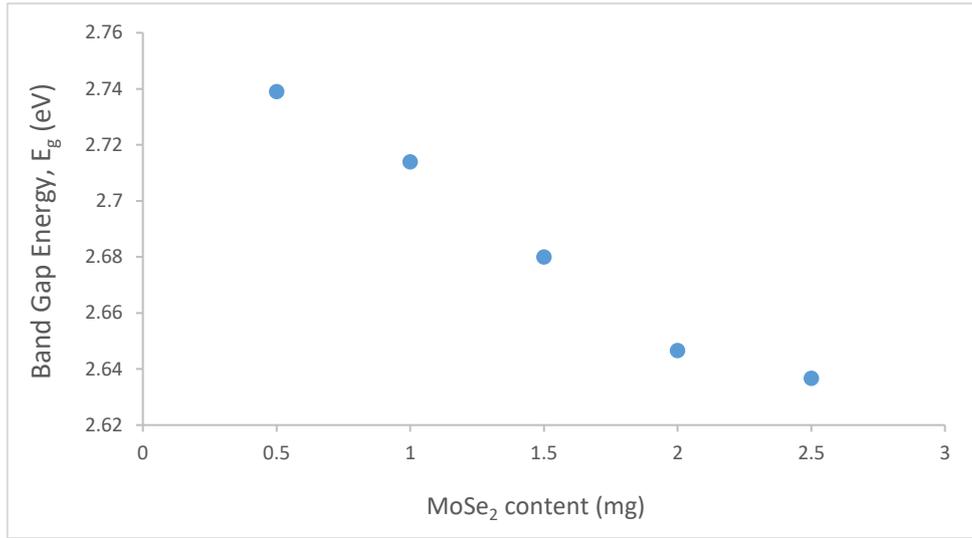

**Figure 6.** MoSe$_2$ content dependent band gap energy of MoSe$_2$-enhanced polyacrylamide composites.

**Table 1.** The transmittance response and calculated band gap energies of all gels from Tauc's plot.

| MoSe$_2$ amount (mg) in MoSe$_2$ + PAAm gel | Transmittance % at 700 nm of wavelength | Bandgap energy of MoSe$_2$ + PAAm (eV) |
|---|---|---|
| 0.5 | 84.6 | 2.739 |
| 1 | 68.8 | 2.714 |
| 1.5 | 53.6 | 2.680 |
| 2 | 51.9 | 2.646 |
| 2.5 | 20.6 | 2.636 |

*3.3. Extinction coefficient and refractive index determination of composite gels*

Extinction coefficient for each composite depending on the wavelength is calculated with the formula using defined absorption coefficient values for each photon energy:

$$\alpha(h\nu) = \frac{4\pi\kappa}{\lambda} \quad (3)$$

where $\kappa$ is the extinction coefficient. Figure 7 shows the variation of the extinction coefficient (obtained using Eq.3) with the wavelength for all composite gels. The $\kappa$ values for all MoSe$_2$ amounts increase with wavelength until they reach to the wavelength around 350 nm in the visible region, then decreases with a further increase in wavelength. For the constant wavelength, for instance at 350 nm, the extinction coefficient values of composite gels were increased because the vacations in the network filled with MoSe$_2$ and reached its maximum value for 2.5 mg of MoSe$_2$.

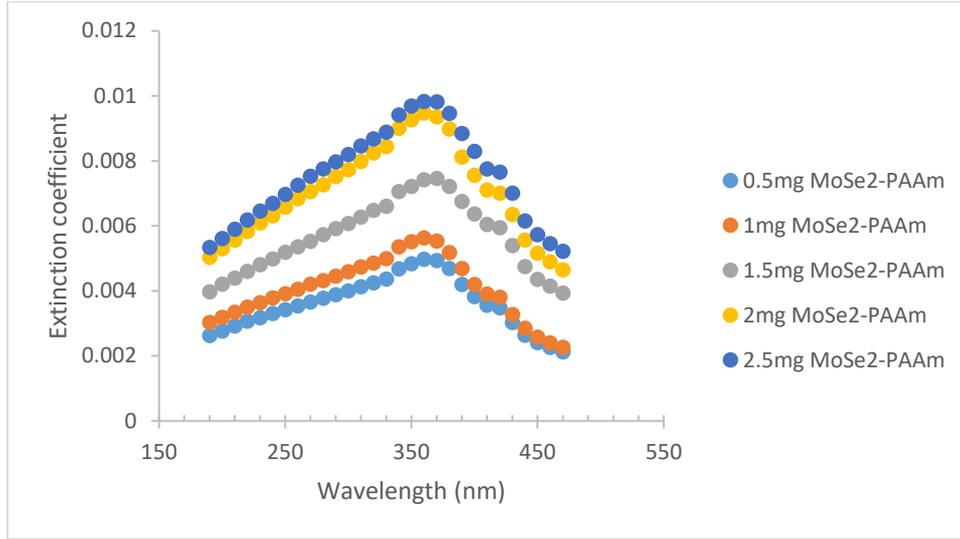

**Figure 7**. Extinction Coefficient of MoSe$_2$-enhanced polyacrylamide composite gels vs wavelength graphs.

Extinction coefficient is actually the component of the frequency-dependent complex refractive index which describes the velocity of propagation of an electromagnetic wave through a solid. The complex refractive index can be given by the following formula:

$$N = n - i\kappa \tag{5}$$

where the real part, known as refractive index, $n$ is related to the velocity, and $\kappa$, the extinction coefficient is related to the decay, or damping of the oscillation amplitude of the incident electric field [32]. Using Fresnel formulas, it is possible to express the refractive index of the material with the following modeling [33]:

$$n(\lambda) = \sqrt{\frac{4R}{(1-R)^2} - k^2} + \left(\frac{1+R}{1-R}\right) \tag{6}$$

Figure 8 represents the refractive index of different composite gels (calculated using Eq.6) versus the wavelength of incident electromagnetic radiation. For all composite gels, refractive index curve has the same trend as decreasing with wavelength. The composite prepared by 2.5 mg MoSe$_2$- PAAm possessed a refractive index of 1.67, but the refractive index of 2.5 mg MoSe$_2$- PAAm composite decreased to 1.60 at wavelength of 500 nm. The refractive index is crucial for purposes of integrated optical devices, such as modulators, filters, and switches [34-38]. Along with the optical dielectric constant, it is regarded as a significant characteristic in the design of new composites for various optoelectronic applications [33]. Ultimately, the new composite obtained has desired band gap energy value and can be actively used in many semiconductor applications [35-38]. Considering the improvements in its mechanical properties, it is predicted that it can also be used as a flexible gel [39]. In addition, the ability to control the refractive index value is promising [40] for its evaluation in optoelectronic applications.

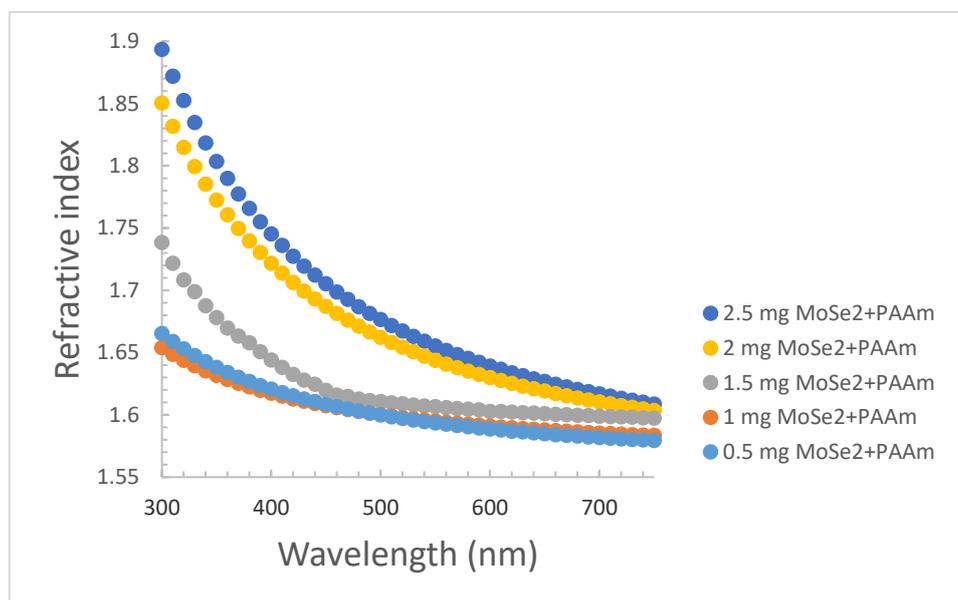

**Figure 8.** Refractive index of composite gels vs wavelength graphs.

## 4. Conclusion

Characteristics of inorganic material enhanced polymer composites continue to be an attractive area of interest for materials scientists. This study aims to uncover the relationship between the tunable band gap energy, refractive index of MoSe$_2$ enhanced polyacrylamide composite gels and their microstructural properties. MoSe$_2$ doped PAAm composites were prepared using free-radical polymerization. The optical properties of MoSe$_2$ doped PAAm were investigated. Increment of MoSe$_2$ amount causes the decrease in the compactness of the composites's macrogel structure.

- It has been postulated that the optical properties (transmission, absorbance, extinction coefficient, refractive index) of MoSe$_2$ enhanced polyacrylamide composites are changing by the MoSe$_2$ content.
- It is obvious to see that the band gap energy decreases with increasing the values of MoSe$_2$ content and then increases with further increases. The calculated band gap energies of the composites are in harmony with the literature. The properties of the final gels are highly dependent on the amount of MoSe$_2$ present during the formation of gels. In this case, it can be concluded that composite gels are better semiconductors regarding the conductivity if the correlation between conductivity and band gap energy is taken into account.

- The band gap energies and refractive index of the composites are found as in agreement with the literature. Furthermore, it is not complicated to obtain MoSe$_2$-enhanced polyacrylamide composites with tunable refractive index and band gap energy.

**Acknowledgments**

The authors would like to thank Prof. Nihat Berker for providing this collaborative working environment and Prof. Sondan Durukanoğlu Feyiz for her support in establishing the Materials Design and Innovation Lab where all the experiments and characterizations of this study were performed.